\documentclass[11pt]{article}

\usepackage[iso-8859-7]{inputenc}
\usepackage{epsfig}
\usepackage{graphicx}

\usepackage{amsfonts}
\usepackage{amssymb}
\usepackage{amsthm}
\usepackage{amsmath}
\usepackage{multirow}

\usepackage{cite}

\newcommand{\newc}{\newcommand}
\newc{\ra}{\rightarrow}
\newc{\lra}{\leftrightarrow}
\newc{\be}{\begin{equation}}
\newc{\ee}{\end{equation}}
\newc{\ba}{\begin{eqnarray}}
\newc{\ea}{\end{eqnarray}}
\newc{\n}{\nu}
\newc{\D}{\Delta}

\newc{\nn}{\nonumber}
\newcommand{\ga}{\alpha}

\begin{document}
\thispagestyle{empty}

\vskip 2truecm
\vspace*{3cm}
\begin{center}
{\large{\bf
Gauge coupling flux thresholds, exotic matter and the unification scale in {\cal F}-$SU(5)$ GUT}}

\vspace*{1cm}
{\bf G. K. Leontaris$^{1,2}$, N.D. Tracas$^3$}\\
$^1$Department of Physics, CERN Theory Division, CH-1211, Geneva 23, Switzerland\\
$^2$Theoretical Physics Division, Ioannina University, GR-45110 Ioannina, Greece\\
$^3$Physics Department, National Technical University, 157 73 Athens, Greece
\end{center}

\vspace*{1cm}
\begin{center}
{\bf Abstract}
\end{center}

\noindent
We explore the gauge coupling relations and the unification scale in {\cal F}-theory $SU(5)$ GUT
broken down to the Standard Model by an internal $U(1)_Y$ gauge flux. We consider  variants with
exotic matter representations which may appear in these constructions and investigate their r\^ole in
the effective field theory model. We make a detailed investigation on the conditions imposed on the
extraneous matter to raise the unification scale and make the color triplets heavy in order to avoid
fast proton decay. We also discuss in brief the implications on the gaugino masses.

\vfill
\newpage

\section{Introducation}

Recent activity on model building in the context of F-theory has received considerable
attention~\cite{Donagi:2008ca,Beasley:2008dc,Donagi:2008kj,Hayashi:2008ba,Beasley:2008kw}\footnote{For
related work see also~\cite{{Marsano:2008jq,Jiang:2008yf,Bourjaily:2009vf,Font:2008id,Andreas:2009uf,Blumenhagen:2009gk,Palti:2009,Chen:2009me,Marchesano:2009rz}}}.
In this set up, gauge symmetries  accommodating successful grand unified
theories (GUTs) are naturally realized on seven branes wrapping appropriate compact surfaces.
As in the case of the heterotic string, there are no Higgs fields in the adjoint, and thus
ordinary GUT symmetries cannot break with the usual Higgs mechanism. In the context of
heterotic string, this inconvenience  spawned new ideas~\cite{Antoniadis:1987dx,Antoniadis:1988cm}
of replacing unified GUTs with modified alternative groups which dispense with the use of
the adjoint representation to break down to the Standard  Model (SM) gauge symmetry.
The advantage of F-constructions compared to the heterotic case, is that the gauge group
can break by turning on suitable field configurations on the compact surface $S$ in
a subgroup of the  GUT symmetry supported by the seven brane. Suppose for example
that the seven brane supports an $SU(5)$ gauge symmetry which, as is well known, is
the smallest unified gauge group accommodating the three gauge groups of the Standard Model.
In this case, as described for example in~\cite{Beasley:2008dc} we can turn on a $U(1)_Y$
internal flux  which breaks the $SU(5)$ gauge group down to the SM gauge symmetry.
There are various potential problems in this approach.
One important issue that arises in this scenario concerns the gauge coupling unification
at the string  scale. We know that  the renormalization group (RG) running
of the  minimal supersymmetric SM gauge couplings is consistent with
the embedding of the three gauge factors in a unified gauge group at around
$M_U\sim 10^{16}$ GeV. This value of $M_U$ is considerably smaller than the scale of the
heterotic string and the Planck mass $M_{Pl}$.  Thus, if one wishes to maintain the idea
of unification of SM gauge couplings in a single gauge group,
it is desirable to work out cases where these two scales decouple and
$M_U/M_{Pl}$ is a small number.
It has been argued~\cite{Beasley:2008dc}, that this can happen in cases where
the seven brane realizing the GUT gauge symmetry wraps a del Pezzo surface.

Another  issue in these constructions related to the mechanism employed to
break the GUT   symmetry is the splitting~\cite{Donagi:2008kj, Blumenhagen:2008aw}
of the values of the three gauge coupling constants
when turning on a flux along the $U(1)_Y$. This splitting  cannot be compatible
with the unification scenario at a scale $M_U\sim 10^{16}$ GeV, at least not in
the context of the MSSM. Nevertheless, the accommodation of the SM gauge symmetry
into a unified group usually comes at the cost of extra matter at scales below $M_U$.
For example, even in the simplest $SU(5)$ embedding of the SM gauge symmetry, the Higgs
doublets are incorporated into complete $SU(5)$ fiveplets together with color triplets
which mediate proton decay.   Additional bulk matter may also be present at scales
below the string scale.  The problem of unification in F-$SU(5)$ is therefore more complicated
than in the minimal GUT scenario~\cite{Hisano:1992mh,Dienes:1996du,Kehagias:2005vz,Ross:2007az}
and a fresh look  into the r\^ole of the extra matter
representations in conjunction with the new gauge coupling relations implied by the fluxes is needed.
It is the purpose of this letter to clarify some of these issues. In section 2 we give a short
description of the F-$SU(5)$ GUT and  its representation content. We discuss the proton
decay  and other potential problems caused by the possible appearance of exotic states.
In section 3 we perform a detailed RG analysis and discuss the
correlation of the exotic matter states and  the unification scale. In section 4 we present
our conclusions.

\section{{\cal F}-$SU(5)$ }

We start with a short description of the salient features of
F-theory model building following the work
of~\cite{Beasley:2008dc,Beasley:2008kw}. In F-theory a gauge
symmetry $G_S$ is supported on seven branes wrapping a del Pezzo
surface S on the internal manifold.  In this set-up, massless
spectrum arises when a non-trivial background field configuration on
$S$ obtains a value along some subgroup $H_S$ of
$G_S\supset\Gamma_S\times H_S$.  The spectrum is found in
representations which arise from the decomposition of  the adjoint
of $G_S$ under $\Gamma_S\times H_S$
\be
{\rm adj}(G_S)=\bigoplus \tau_j\otimes T_j
\ee
The net number of chiral  minus anti-chiral
states is given in terms of a topological index
formula~\cite{Beasley:2008dc,Donagi:2008kj},
\[
n_{\tau}-n_{\tau^*}=-\int_Sc_1({\cal T})c_1(S)=\chi(S,{\cal T}_j^*)-\chi(S,{\cal T}_j)\nn
\]
where $\tau^*$ is the dual representation of $\tau$, ${\cal T}$ is the bundle transforming in the
representation $T$ and $\chi$ is the Euler character.

In a more general background containing intersecting branes,  chiral matter appears
on Riemann surfaces $\Sigma_i$ which are located at the intersection loci of the compact surface S with
other in general non-compact surfaces $S_1, S_2,\cdots$. These chiral states appear in bifundamental
 representations in close analogy to the case of intersecting D-brane models. Along the intersections the rank of the singularity increases. Designating $G_S$ the gauge group on the surface $S$ and $G_{S_i}$ that associated with  $S_i$, the gauge group on $\Sigma_i$ is enhanced to $G_{\Sigma_i}\supset G_S\times G_{S_i}$ whose   the adjoint in general  decomposes as
\be
{\rm adj}(G_{\Sigma_i})= {\rm ad}(G_{S})\oplus {\rm ad}(G_{S_i})\oplus (\oplus_j U_j\otimes {(U_i)}_j)
\ee
with $U_j,{(U_i)}_j$ being the irreducible representations of $G_S, G_{S_i}$.  In the simple case of $G_{S}=SU(n)$,
$G_{S_i}=SU(m)$,  and $G_{\Sigma_i}=SU(n+m)$ for example, the chiral ${\cal N}=1$ multiplet is
the bifundamental $(n,\bar m)$.

We assume that a non-trivial background gauge field configuration acquires a value in a subgroup $H_S\subset G_S$ and similarly in $H_{S_i}\subset G_{S_i}$. If $G_S\supset\Gamma_S\times H_S$ and
 $G_{S_i}\supset\Gamma_{S_1}\times H_{S_i}$, with $\Gamma_{S,S_i}$ being the corresponding maximal
 $G_{S,S_i}$ subgroups,  the $G_S\times G_{S_i}$ symmetry breaks to the commutant
group $\Gamma =\Gamma_S\times \Gamma_{S_i}$. Denoting also $H=H_S\times H_{S_i}$, the decomposition
of $U\otimes {U_i}$ into irreducible representations of  $\Gamma\times H$ give
\be
U\otimes {U_i} = \oplus_j (r_j,R_j)
\ee
The net number of chiral fermions $n_{r_j}-n_{r^*_j}$ in a specific representation is
\be
n_{r_j}-n_{r^*_j} = \chi(\Sigma_i,K_{\Sigma_i}^{1/2}\otimes V_j)
\ee
$K_{\Sigma_i}^{1/2}$ being the spin bundle over $\Sigma_i$ and $V_j$ that
of $R_j$.
In the case of an algebraic curve $\Sigma_i$ the Euler character is written
 as a function of the genus of the Riemann surface $\Sigma_i$ and the first Chern
 class~\cite{Beasley:2008dc}.

Having described the basic features of the F-theory constructions, we discuss now
the minimal  unified gauge group  that this scenario can be
realized~\cite{Beasley:2008dc}-\cite{Bourjaily:2009vf}.
 In order to obtain a viable effective low energy model, the seven brane wrapping the
 del Pezzo surface $S$ must support at least an $SU(5)$ gauge group which is the minimal GUT
 containing  the SM gauge symmetry. The symmetry breaking down to the SM
cannot occur via the adjoint Higgs field since the GUT surface does not admit adjoint scalars.
The advantage of constructing $SU(5)$ in F-theory is that it  is possible to turn on a non-trivial
$U(1)_Y$ flux  which breaks $SU(5)$ to the Standard Model gauge group.
This flux is considered as a connection on a line bundle ${\cal L}_Y$ and taken to be localized,
so that it can be non-trivial
on the seven brane wrapping $S$ but trivial in the base of the F-theory compactification,
 thus the final group can be $SU(3)\times SU(2)\times U(1)_Y$.
The flux also determines the matter context of the  low energy effective field theory model.
 In particular, in F-$SU(5)$
the possible representations with their decompositions under $SU(3)\times SU(2)\times U(1)$
are as follows
\begin{align}
5&\ra ( 3,1)_{-2}+(1,2)_{3}\\
\bar 5&\ra (\bar 3,1)_2+(1,2)_{-3}\\
10&\ra(\bar 3,1)_{-4}+(3,2)_1+(1,1)_6\\
\overline{10}&\ra( 3,1)_{4}+(\bar 3,2)_{-1}+(1,1)_{-6}\\
24&\ra(8,1)_0+(1,3)_{0}+(1,1)_0+(3,2)_{-5}+(\bar 3,2)_5\label{24bu}
\end{align}
The $U(1)$ normalization to obtain the correct charges  is $Y=\frac 16 Q_{U(1)}$, with
the electric charge given by $Q=T_3+Y$.
The fermion families belong to $10_F$ and $\bar 5_f$,
\begin{align}
10_F&\ra u^c+Q+e^c\\
\bar 5_f&\ra  d^c+\ell
\end{align}
and the SM Higgs fields to $5_h+\bar 5_{\bar h}$
\begin{align}
 5_h&\ra  D_h+h_u\\
\bar 5_{\bar h} &\ra \bar D_h+h_d
\end{align}
The masses arise from the couplings
\begin{align}
10_F\,10_F\,5_h&\ra Q\,u^c\,h_u+u^c\,e^c\,D_h+Q\,Q\,D_h\label{upY}\\
10_F\,\bar 5_f\,\bar 5_h&\ra Q\,d^c\,h_d+e^c\,\ell\,h_d+u^c\,d^c\,\bar D_h+Q\,\ell\,\bar D_h
\label{doY}
\end{align}
Clearly, the necessary fermion mass terms are accompanied by dangerous trilinear terms
which combine to the well known dimension five operators causing proton decay. We will
deal with this issue in the context of F-$SU(5)$ in the next subsection. We note in passing that
there could be also R-parity violating terms
\begin{align}
10_F\,\bar 5_f\,\bar 5_f&\ra Q\,d^c\,\ell+e^c\,\ell\,\ell+u^c\,d^c\,d^c
\end{align}
which must be absent due to some ($U(1)_{PQ}$-type) symmetry, or they have to be highly
suppressed to avoid rapid proton decay.

\subsection{Exotics and Proton decay}

From the decomposition of the available $SU(5)$ representations in F-$SU(5)$, we see that in
addition to the minimal SM spectrum there are also extra states which in principle could appear
in the light spectrum. Below we discuss in brief their consequences and how they can be removed
from the low energy effective theory.

Decomposing  the adjoint of $SU(5)$ under the SM gauge symmetry,
one finds the representations $Q^\prime =(3,\bar 2)_{-5}$  and $\bar Q^\prime=(\bar 3,2)_5$
which carry non-trivial
$U(1)$ charges. Usually, these bulk exotic states appear in pairs because of well defined
transformation properties under discrete symmetries. The $Q^\prime$ field has the exotic charges
\be
Q^\prime\equiv (3,\bar 2)_{-5}=\left(
\begin{array}{c}
                       \xi_{-\frac{1}{3}} \\
                       \zeta_{-\frac{4}{3}} \\
                     \end{array}\right)
\ee
whilst $\bar Q^\prime$ has their conjugates.

 The appearance of $Q^\prime, \bar Q^\prime$ in the  spectrum of the effective theory has two undesired
consequences.
 First, as we will see in detail in the next section, they modify the beta functions of the
SM gauge couplings and as a result they lower the unification scale. Second, they can form
couplings of the type $S\Sigma\Sigma$ with Standard Model matter fields. The following
couplings originate involving bulk fields
\begin{align}
24_{S}\cdot 5_{\Sigma}\cdot\bar 5_{\Sigma}&\ra (3,\bar 2)_{-5}(1,2)_3(\bar 3,1)_2\ra Q^\prime h_u \bar D_h,\, Q^\prime h_u d^c\nn\\
               &+ (\bar 3,2)_{5}(1,2)_{-3}(3,1)_{-2}\ra \bar Q^\prime h_d  D_h,\,\bar Q^\prime \ell  D_h\nn
\end{align}
These include terms of the form   $ \zeta\,h^+_u\,d^c+\xi\,h^0_u\,d^c+\cdots$, thus the
$\zeta$-field  decays to $d^c$ and $\xi$-field couples through a mass term to $d^c$. The following mass terms can exist
\be
h_d^0 d\,d^c+h^0_u \xi d^c+M_{KK}\bar \xi\,\xi
\ee
which imply an unacceptable mixing between the ordinary down quark  mass $3\times 3$ matrix
$m_d$ and the exotic(s) state $\xi$:
\be
m_{down}\sim\left(\begin{array}{cc}m_d&0\\m_u&M_{KK}\end{array}\right)
\ee
This problem  suggests that either the extra fields $Q^\prime,\bar Q^\prime$ should be absent, or
that the couplings of type ($S\Sigma\Sigma$) should be zero.\footnote{We note in passing that in the presence
of additional states originating from $10,\overline{10}$ vector
fields more couplings of the following form might be present
\be
24\cdot 10\cdot\overline{10} \ra Q^\prime\,Q\,\bar u^c_H+\,Q^\prime\bar Q_H\,e^c
+\bar Q^\prime\bar Q_Hu^c+\bar Q^\prime Q\bar e^c_H\nn
\ee
}

A second drawback in $SU(5)$ models comes from the unsolicited color triplets $D_h,\bar D_h$
 which are always present in the GUT spectrum.
They are constituents of the same $5,\bar 5$ multiplets  where the Higgs doublets -needed to
break the SM gauge symmetry- are found. It is well known that in the minimal $SU(5)$ there
are dimension
five operators generated by $D_h,\bar D_h$  as well as dimension  six operators
from diagrams mediated
by  gauge bosons, which both induce proton decay.  Among them, the most dangerous ones are those
which are constructed by the exchange of the color scalar triplets discussed above.
A relevant graph generates in the superpotential the effective operator
\cite{Weinberg:1981wj,Sakai:1981pk,Babu:1998wi,Murayama:2001ur,Dermisek:2000hr},
\be
{\cal W}_5\sim\frac{\lambda_1^i\lambda_2^k}{M_{eff}}V_{jk}^* Q_iQ_iQ_j\ell_k
\label{pdo5}
\ee
where the coupling $\lambda_1$ is related to the corresponding up-quark Yukawa coupling (see
(\ref{upY}) ), while $\lambda_2$ has a common origin with the Yukawa coupling for the down quarks (see (\ref{doY})).
$V_{jk}^*$ is the CKM mixing element and $M_{eff}$ is an effective scale related to the mass of the color triplet
and their exact relation is determined when the entire color triplet mass matrix is
 specified\cite{Babu:1998wi,Murayama:2001ur,Dermisek:2000hr}.

The operator (\ref{pdo5}), when dressed with charged-wino and/or higgsino
leads to the most dominant decay $p\ra K\bar \nu$.
Therefore these triplets must be heavy enough in order to imply a proton decay rate
consistent with the present bounds. According to experimental bounds
put by KamioKande ($\tau_p\ge 6.7\times 10^{32}$ years), even for relatively small Yukawa
couplings $\lambda_{1,2}<1$ the  triplet mass must be at least
heavier than $\sim 7\times 10^{16}{\rm GeV}$.
If the GUT breaking scale is that of the minimal supersymmetric $SU(5)$ model,
$M_G\sim 2\times 10^{16}$ GeV,  the problem is clear, since the triplet mass
is of the order of $M_G$. On way  to evade this
shortcoming in F-theory, is to assume that the two Higgs fields are localized
at different curves and generating heavy mass terms with another  triplet pair
so that the effective operator (\ref{pdo5}) can be avoided\cite{Beasley:2008kw}.
We notice however that this color triplet arrangement may also impose restrictions
or possible zero entries (texture-zeros) on  the mass matrices.

In the following we wish to elaborate on a complementary solution to this problem.
In particular we will examine the possibility  of raising the $SU(5)$ GUT scale
at least at the level of the heterotic string scale\footnote{In
the cases of  $M_U$ scales comparable to $M_{Pl}$ one may relax  the assumption
of decoupling discussed in the introduction.} $M_U\sim 2\times 10^{17}$ GeV so that
the mass of the triplets can be
heavy enough.  We may consider this possibility in conjunction with the fact
that  in F-theory constructions it is possible that additional matter representations
arise as vector like pairs in the bulk or on some matter curves $\Sigma_i$.
In this case one can take advantage of these extra matter and proceed to suitable
modifications~\cite{Babu:1994dq} of the doublet-triplet splitting problem in order
to avoid rapid proton decay.

 The possible vector like multiplets available in these constructions
descend from the representations $10+\overline{10}$ and $5+\overline{5}$
and thus have the quantum numbers of ordinary SM states.
When these latter representations form complete $SU(5)$ multiplets they
do not modify  the value of the GUT scale.
However,  individual pairs of them do have non-vanishing effects on the
GUT scale. In a viable model  is expected of course that at some
scale below the unification they will receive masses and decouple from the light
spectrum. Depending on the particular combinations that appear
 in the spectrum, these states can decrease  or increase the unification scale. In a
 bottom-up approach, one may determine the appropriate spectrum to obtain
 a `unification' point at energies high enough to avoid rapid proton decay and then
 determine the admissible line bundle configurations that lead to the desired
 zero mode context of the effective theory.

\section{Gauge coupling relations and the GUT scale}

The issue of the Standard Model gauge couplings running and their integration  into an
${\cal N}=1$  Supersymmetric  $SU(5)$  GUT in the context of F-theory was considered in
\cite{Blumenhagen:2008aw}.  In this work it was observed that  the $U(1)_Y$ flux turned
 on to break the $SU(5)$
symmetry leads to a splitting of gauge couplings at the unification scale.
In the presence of the line bundles discussed above, the following  gauge coupling relations
are derived at the string scale\cite{Blumenhagen:2008aw}
\be\label{gcMU}
\begin{split}
\frac{1}{a_3(M_G)}&=\frac{1}{a_G}-y
\\
\frac{1}{a_2(M_G)}&=\frac{1}{a_G}-y+x
\\
\frac{1}{a_1(M_G)}&=\frac{1}{a_G}-y+\frac 35 x
\end{split}
\ee
where $x=- \frac 12 S\int c_1^2({\cal L}_Y)$ and  $y=\frac 12 S\int c_1^2({\cal L}_a) $
associated with a non-trivial line bundle ${\cal L}_a$ and $S=e^{-\phi}+\imath C_0$ the
axion-dilaton field as discussed in~\cite{Blumenhagen:2008aw}.
Then
\begin{align}
\frac{1}{a_2(M_G)}-\frac{1}{a_3(M_G)}&=x\\
\frac{1}{a_1(M_G)}-\frac{1}{a_3(M_G)}&=\frac 35 \,x
\end{align}
so that the following relation is found at the unification
scale\cite{Blumenhagen:2008aw}
\be
\frac 53 \,\frac{1}{a_1(M_G)}=\frac{1}{a_2(M_G)}+\frac 23 \frac{1}{a_3(M_G)}\label{SR}
\ee
In addition, taking into account (\ref{gcMU}) and the fact that $-\frac 12 \int c_1^2({\cal L}_Y)=1>0$,
the following hierarchy of the couplings holds at $M_G$:
\be\label{ineq}
a_3(M_G)\ge a_1(M_G)\ge a_2(M_G)
\ee
where the equalities hold in the case of $x=0$. This limiting case can be reached by
twisting ${\cal L}_a=\imath^*(L_a)$ appropriately by a trivial line
bundle~\cite{Donagi:2008kj,Palti:2009,Blumenhagen:2008aw}.

In view of (\ref{SR}) and (\ref{ineq}) non-standard GUT relations, in what follows, we will investigate
in some detail the issue of unification in the presence of extra matter  threshold
effects.

\subsection{The case of the color triplets $D_h,\bar D_h$}

We consider first the simple case where only the triplet pairs appear in the spectrum
below the $SU(5)$-GUT breaking  scale $M_G$. We
assume that at some scale $M_X<M_G$  the extra triplet pairs $D_h,\bar D_h$  decouple
and only the MSSM spectrum remains massless for scales $\mu<M_X$.
The low energy values of the  gauge couplings are then given by the evolution equations
\be\label{Brun}
\frac{1}{a_i(M_Z)} = \frac{1}{a_{i}(M_G)}+\frac{b_i^x}{2\pi}\,\ln\frac{M_G}{M_X}+
\frac{b_i}{2\pi}\,\ln\frac{M_X}{M_Z}
\ee
where $b_i^x, (b_i)$ are the beta-functions above (below) the scale $M_X$ and $a_{i}(M_G)$
given by (\ref{gcMU}).

Using the GUT relation (\ref{SR}) one arrives at the following equation\cite{Blumenhagen:2008aw}
\be
\label{D}
\left[5 (b_1^x-b_1)- 2(b_3^x-b_3)\right]\,\ln\left(\frac{M_G}{M_X}\right)=0
\ee
For $n$ triplet pairs the differences $b_1^x-b_1=\frac 15 \cdot 2 n$ and $b_3^x-b_3=\frac 12 \cdot 2n$.
Therefore, as it was observed\cite{Blumenhagen:2008aw}, the expression with the $\beta$-functions
 in (\ref{D}) vanishes and the equation is satisfied for any number of triplets irrespectively
of the scale $M_X$ they become massive. Thus,  for a given value of $x$-shift we can choose the
scale $M_X$ where the triplets  decouple to modify appropriately the gauge coupling running
so that (\ref{SR}) and (\ref{ineq}) are fulfilled.

\begin{figure}[!b]
\centering
\includegraphics[scale=.6,angle=0]{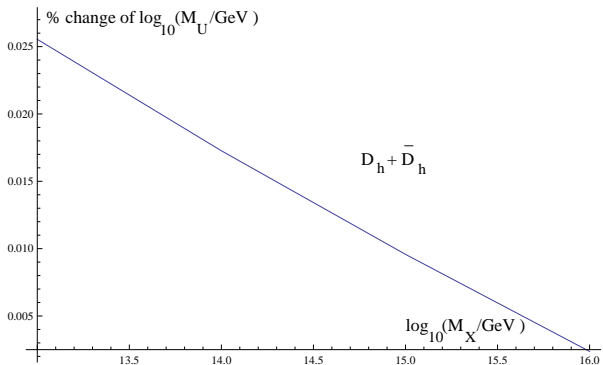}
\caption{Percentage correction, between 2- and 1-loop running, on the $SU(5)$ GUT scale $M_U$ as a
function of the decoupling scale of a color triplet pair $D_h+\bar
D_h$.} \label{D_2LOOP}
\end{figure}

This exact $M_X$-scale independence at the one-loop level, is of course spoiled
when two-loop contributions to the beta-functions are taken into account. To estimate
the effect, we run the renormalization group equations for the gauge couplings at
two loop, including one-loop contributions from the top quark coupling.
In Fig.(\ref{D_2LOOP}) we plot the percentage change of the unification scale
as a function of the scale where the triplets acquire a mass. We notice only a marginal change
from the ``no-extra-matter'' scenario.
We also mention that in 1-loop running the scale where (\ref{SR}) is satisfied, with no extra matter at all,
is $2.15\times 10^{16}$ GeV while in 2-loop running this scale is slightly higher $2.21\times 10^{16}$ GeV.

\subsection{The general case }

We assume next the more general situation where various types of extra states
 remain down to a scale $M_X<M_U$ after the breaking of the GUT symmetry $M_U$.
When we take into account  the contributions of complete $SU(5)$
($10/\overline{10},5/\overline{5}$) vector like multiplets, the unification
point does not  change, irrespectively of the scale these multiplets
become massive. However, incomplete multiplets  may originate from
the bulk surface $S$ as vector like pairs. In fact, for the $SU(5)$
case, the exotic superfield pair  $(3,\bar 2)_{-5}+(\bar 3, 2)_5$
descends from the decomposition of the adjoint representation. It is
also possible  that after the $SU(5)$ breaking other type of extra
vector like states to appear in the spectrum along the various
matter curves $\Sigma_i$.
In Table (\ref{betacon}) we summarize the
contributions of the various SM representations to the beta
functions.
\begin{table}[!b]
\centering
\begin{tabular}{c|cccc}
   &   $b_Y$  &   $b_2$  &  $b_3$  &  $\delta\beta$\\
\hline
$Q$ &$1/6$ &$  3/2 $&$ 1    $&$ -2  $ \\
$u$ &$4/3$ &$      $&$ 1/2  $&$  1  $ \\
$d$ &$1/3$ &$      $&$ 1/2  $&$  0  $ \\
$L$ &$1/2$ &$  1/2 $&$      $&$  0  $ \\
$e$ &$ 1       $ &$      $&$      $&$  1  $ \\
$H$ &$3/10$ &$ 1/2 $&$      $&$  0  $ \\
$Q^\prime$ &$1/2$ &$1/2 $&$ 0 $&$0  $ \\
\end{tabular}
\caption{The contributions of the various extra  states to the beta functions
and the combination $\delta\beta=\delta (b_Y-b_2-(2/3)b_3)$.  }
\label{betacon}
\end{table}
To investigate their r\^ole in the determination of the string scale, we proceed to a more
general analysis of the RGEs at the one and two-loop level. To get an intuition of the
specific contribution of each of the states presented in Table (\ref{betacon})  we start first with
the exploration at the one-loop level where the analytic formulae are easy to handle. Thus,
combining  equations  (\ref{SR})  and  (\ref{Brun}), we obtain
\be
M_U =  e^{\frac{2\pi}{\beta {\cal A}}\rho}\,\left(\frac{M_X}{M_Z}\right)^{1-\rho}
M_Z\label{M_U}
\ee
with
\be
\rho  = \frac{\beta}{\beta_x}
\ee
and $\beta,\beta_x$ are the beta-functions combinations
\begin{align}
\beta_x&=b_Y^x-b_2^x-\frac 23b_3^x\label{betax}\\
\beta&=b_Y-b_2-\frac 23b_3\label{beta0}
\end{align}
while ${\cal A}$ is a function of the experimentally known low energy values of the
SM gauge coupling constants
\be
\frac{1}{\cal A} = \frac 53 \,\frac{1}{a_1(M_Z)}-\frac{1}{a_2(M_Z)}-\frac 23 \frac{1}{a_3(M_Z)}
\ee
For $\rho=1$, i.e., when there are no extra contributions in the beta functions
 or -more precisely- when the extra content contribution to the specific combination of the beta
 functions vanishes, we obtain
the previous value of the GUT scale $M_U=M_G=e^{\frac{2\pi}{\beta {\cal A}}}M_Z\approx 2.1\times 10^{16}$ GeV.
We can substitute this into (\ref{M_U}) to obtain a more illuminating  relation between the `old'
$M_G$ and `new' unification scale $M_U$
\be
\frac{M_U}{M_G} = \left(\frac{M_X}{M_G}\right)^{1-\rho}\label{MUG}
\ee
Furthermore, as can be seen from Table (\ref{betacon}) the  contributions of the triplets to
one-loop $\beta_x$ beta-function combination are zero.
Therefore the scale $M_D$ at which they become massive does not affect
the scale $M_U$ at which (\ref{SR}) is fulfilled and of course,
does not necessarily coincide with $M_X$ where the remaining extra states decouple.
Now we can classify easily all the cases.
\begin{itemize}
\item
If the contribution of extraneous matter in the beta functions  combination $\beta_x$ is zero,
then  $\beta_x=\beta$, or $\rho=1$ and we have no dependence on the $M_X$ scale.
In this case the  exponent in (\ref{MUG}) is zero and then we obtain the old GUT scale $M_U=M_G$ .
\item
If $0<\rho<1$, (i.e., $\beta_x>\beta$),  consistency of the hierarchy of scales
requires that we take  the extra matter to be massive at scales%
\footnote{
If we take $M_X>M_G$, we see from $\frac{M_U}{M_X}=\left( \frac{M_G}{M_X}\right)^{\rho}$ that
this would imply $M_U<M_X$ which is unacceptable.}  $M_X<M_G$, the scale $M_U$ is suppressed.
\item
Finally, for $\rho> 1$ the power $1-\rho$ of the same factor is negative and
we can take only  $M_X< M_G$,  the $M_U$ scale is enhanced.
\end{itemize}

\begin{figure}[!t]
\centering
\includegraphics[scale=.7,angle=0]{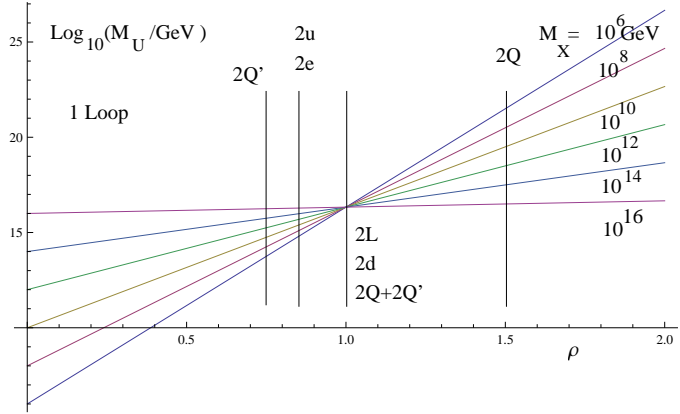}
\caption{The unification scale $M_U$ as a function of $\rho=\beta/\beta_x$ for several energy scales $M_X$
where the extra matter is assumed to contribute  the $\beta$-functions. We also indicate the $\rho$ value
for several specific cases of (extra) matter content. We note that in practice $\rho$ takes only
discrete values and that only a subset of the above cases is compatible
with the gauge coupling inequality (\ref{ineq}). See text for details. }
\label{rho}
\end{figure}

To get an insight of the impact of the extra matter on the determination of  the
 scale $M_U$, in Fig.(\ref{rho}) we plot the scale $M_U$ vs a sufficiently wide range of
 $\rho$ values assuming that the extra matter states receive masses at a scale $M_X<M_G$.
  In this graph  each line corresponds to a certain  common (average)
 scale $M_X$ where  the extra matter becomes massive.
 When $\rho=1$, i.e., when there is no extra matter, or more precisely when  there is no
 contribution of the extra matter to the combination (\ref{beta0}), we obtain the
 standard $SU(5)$ GUT scale $M_U\equiv M_G\sim 2\times 10^{16}$ GeV. As expected
 all lines of the graph meet at the same point which is in accordance with the observation above.
 For values $\rho<1$,  the $M_U$  scale decreases while the effect is obviously enhanced for
  lower $M_X$ scales. This happens  when the  extra matter is composed by states
  which assign positive contributions to (\ref{beta0}).  Several vertical lines
  have been drawn on the  graph to show the effect of various individual pairs
on the unification scale. It is observed that the addition of
 an exotic $Q^\prime+\bar Q^\prime$ bulk pair descending from the adjoint reduces the unification
scale to unacceptable values. For this reason as well as the unacceptable mixing
with ordinary matter induced by the couplings (\ref{pdo5})
these exotic states must be eliminated from the spectrum.  To this end, if $L$ is assumed to be the line bundle
on $S$ associated with the breaking of $SU(5)$,  we must impose the condition $\chi(S,L^{\pm 5})=0$.
\footnote{Of course, if $L$ is a line bundle, $L^5$ cannot also correspond to a root of a Lie Algebra,
however, it was shown that fractional powers of line bundles are also consistent
\cite{Beasley:2008kw}.} In the presence of sufficiently large number of bulk states
the parameter $\rho$ can also attain values beyond the range discussed above and we will see
specific examples in the subsequent analysis.

We have already pointed out that the admissible energy scales with their inextricable exotic matter spectra
are restricted by the inequality (\ref{ineq}) that holds among the gauge couplings' values at the unification scale.
We can rewrite these inequalities in terms of the beta functions
\begin{align}
\delta\left(b_3-\frac 35 b_Y\right)\,\ln\frac{M_U}{M_X}&>+\frac{48}5\,\ln\frac{M_U}{M_Z}-2\pi\left(\frac 35\frac{1}{a_Y}-\frac{1}{a_3}\right)\label{inq1}
\\
\delta\left(\frac 35 b_Y-b_2\right)\,\ln\frac{M_U}{M_X}&>-\frac{28}5\,\ln\frac{M_U}{M_Z}+2\pi\left(\frac 35\frac{1}{a_Y}-\frac{1}{a_2}\right)\label{inq2}
\end{align}
where in the right-hand side we have substituted the numerical values of the MSSM beta functions.
On the left hand side of these constraints
we can express the differences in terms of the integer numbers $n_Q,n_u,n_d,n_L,n_e$ representing the multiplicities of the
extraneous matter introduced in the spectrum above $M_X$.  The $a_{Y,2,3}$  couplings  on the right hand side assume values at $M_Z$,
thus for a desired unification scale $M_U\sim 10^{17}$ GeV, we can turn these inequalities to constraints on the extra matter representations.
Clearly, the above inequalities imply strong restrictions on the exotic matter needed to raise the unification scale
while being  compatible with the flux-modified conditions on the gauge coupling hierarchies at $M_U$.
Expressing the beta-functions in terms of the extra matter multiplicities  the above inequalities can be written
\[
\begin{split}
\left(n_d-2n_e-n_u-n_L+3 n_Q\right)\,\ln\frac{M_U}{M_X}&> 32\,\ln\frac{M_U}{M_Z}-4\pi\left(\frac{\cos^2\theta_W}{a_{em}}-\frac{5}{3a_3}\right)\\
\left(n_d+3n_e+4n_u-n_L-7n_Q\right)\,\ln\frac{M_U}{M_X}&>-{28}\,\ln\frac{M_U}{M_Z}+2\pi\frac{3-8\sin^2\theta_W}{a_{em}}
\end{split}
\]
\begin{figure}[!t]
\centering
\includegraphics[scale=0.7,angle=0]{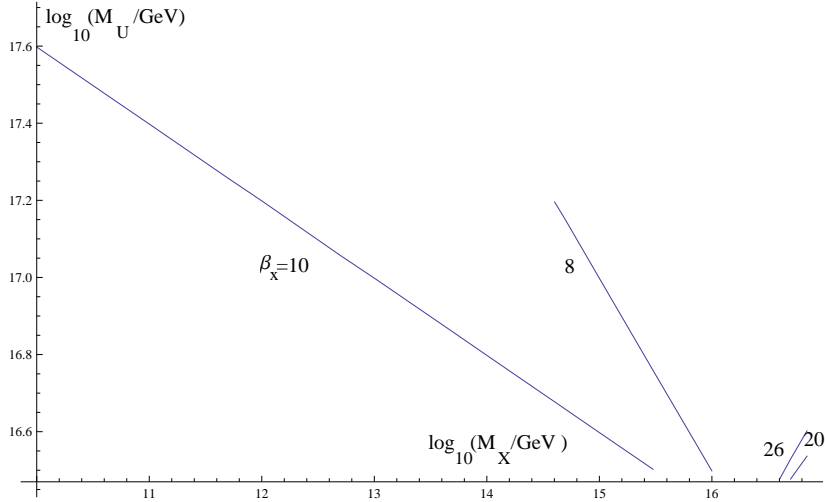}
\caption{The unification scale $M_U$ as a function of the scale $M_X$ (1-loop running) for extra
matter context $\beta_x=10,8,$ and  $\beta_x=20,26$.}
\label{DIOFANDOS}
\end{figure}
Undoubtedly, there are various combinations of extra matter representations satisfying the above conditions.
In the following, we restrict our analysis to cases with minimal numbers of extra matter while
in our calculations we incorporate corrections implied by two-loop contributions to the beta functions.
We will also assume that the triplets become massive at the string scale $M_U$, thus they do not affect
in any way the inequalities or  the  two-loop RG running.

We start with the one-loop analysis and depict our results in Fig.\ref{DIOFANDOS}.
We choose to plot the unification scale $M_U$  as a function of the
 extra matter decoupling scale $M_X$, for various values of the beta function
combination $\beta_x$ given  in (\ref{betax}).

For $\beta_x=10$ we see that for  $M_X$ values in the range $[10^{10}-4\times 10^{14}]$ GeV,  the
corresponding unification scale ranges $$M_U\sim 3\times 10^{16}-3.9\times 10^{17}{\rm GeV}.$$
As far as we keep the extra matter below five pairs, the only allowed combinations of extra matter
consistent with the inequalities, is 2 pairs of Q-like and 3 pairs of u-like particles or 3 pairs
of Q-, 4 pairs of u- and 1 pair of e-like matter (see also Table \ref{exma}).
Note that since we are interested in high values of the unification scale,
 we draw the lines only for scales bigger than $M_U>3\times 10^{16}$ GeV. For the $\beta_x=10$
 case in particular, the line is drawn  down to a  reasonable lower decoupling scale $M_X\sim 10^{10}$ GeV,
 although the conditions (\ref{inq1},\ref{inq2}) would allow even for smaller $M_X$ scales.

\begin{table}[!h]
\centering
\begin{tabular}{c|ccc}
\hline
$\beta_x     $        &  \textrm{extra matter}  & \textrm{min}\,$M_X$(\textrm{GeV})        &   \textrm{max}\,$M_U$(\textrm{GeV})\\
\hline
\multirow{8}{*}{10} &   $4Q,\, 6u  $            & \multirow{2}{*}{$10^{10}$}         & \multirow{2}{*}{$3.96\times 10^{17}$} \\
                    &  $ 6Q,\,8u,\,2e   $       &                                    &                                       \\
\cline{2-4}
                    &  $ 2Q,\,2u   $            & \multirow{6}{*}{$2\times 10^{15}$} & \multirow{6}{*}{$3.18\times 10^{16}$} \\
                    &  $ 4Q,\,4u,\,2e    $      &                                    &                                       \\
                    &  $ 4Q,\,6u,\,4L  $        &                                    &                                       \\
                    &  $ 6Q,\,6u,\,4e     $     &                                    &                                       \\
                    &  $ 6Q,\,8u,\,2L,\,2e $    &                                    &                                       \\
                    &  $ 8Q,\,8u,\,6e$          &                                    &                                       \\
\hline
\multirow{5}{*}{8} & $  6Q,\,8u   $            &$4\times 10^{14}    $                 & $  1.57\times 10^{17}      $            \\
\cline{2-4}
                    & $  4Q,\,4u   $            &\multirow{4}{*}{$10^{16}$} & \multirow{4}{*}{$3.15\times 10^{16}$}          \\
                    & $  6Q,\,6u,\,2e  $        &                           &                                                \\
                    &   $6Q,\,8u,\,2L    $      &                           &                                                \\
                    &   $8Q,\,8u,\,4e  $        &                           &                                                \\
\hline
\end{tabular}
\caption{The predictions for the $SU(5)$ breaking scale and the
corresponding decoupling scale $M_X$ for  two values of the
combination $\beta_x= b_Y^x-b_2^x-(2/3)b_3^x$. Various types of
additional matter result to the same $\beta_x$ and therefore to the
same $M_U$ prediction. }
\label{exma}
\end{table}
As a second possibility, we assume the case of extra matter content which gives  $\beta_x=8$
corresponding to the combination of 3 pairs of Q and 4 pairs of u.
Now consistency  with (\ref{inq1},\ref{inq2})  requires that the extra matter decouples at
scales  $M_X>4\times 10^{14}$ GeV. We observe that this case corresponds to a string scale
as high as $M_U\sim 1.6\times 10^{17}$ GeV.  For higher $M_X$  values more extra matter combinations respect
 the requirements, however the conditions (\ref{inq1},\ref{inq2}) restrict  the string scale to lower
 and lower values.  Because of the $\ln(M_U)$ dependence  on the factor $1-\beta/\beta_x=1-\rho$,
 we  observe that as long as $0<\beta_x<\beta=12$ the slope of the lines are  negative thus the lower
 the $M_X$ decoupling scale, the higher the unification mass $M_U$.
For $\beta_x>\beta=12$ the slope of the lines changes and the opposite is true.

When two-loop corrections are taken into account, the $SU(5)$ GUT scale may attain even higher values.
To show the effect, we pick up the  case $n_Q=4$ and $n_u=6$ of extra matter representations.
We run numerically the coupled RG equations using the SM beta-functions from $M_Z$ to $m_{top}$,
and the MSSM spectrum from $m_{top}$ to $M_X$, while we take into
account the extra matter contributions in the range $M_X-M_U$. In Table \ref{2LR} we present the results for
the unification scale and the values of the gauge couplings $a_i(M_U)$ as they are shifted by the fluxes' threshold
corrections.  From the findings presented in this table we conclude that in the
presence of a rather moderate number of additional matter states it is possible to obtain a unification
scale $M_U$ sufficiently larger than the  ordinary GUT breaking scale of the minimal $SU(5)$.  This
scale is of the order $10^{17}$ GeV and therefore
allows the possibility  to make the colored triplets incorporated in the fiveplets heavy enough in order
to avoid fast proton decay.
 In the last column we present also the corresponding values of the `flux' parameter $x$ which is
intimately  related to the dilaton field. These values imply  dilaton vevs leading to a strong
coupling regime which is the appropriate description for F-theory.
\begin{table}[!h]
\[
\begin{array}{|c|c|c|c|c|c|}
\hline M_X(\textrm{GeV})    & M_U(\textrm{GeV})                & \ga_3(M_U) & (5/3)\ga_1(M_U)&
\ga_2(M_U)& x       \\ \hline 10^{11}& 6.23\times 10^{17} & 0.15680
& 0.15097        & 0.14730   & 0.41035 \\ \hline 10^{12}& 2.90\times
10^{17} & 0.09927    & 0.09734        & 0.09609   & 0.33346 \\
\hline 10^{13}& 1.51\times 10^{17} & 0.07429    & 0.07354        &
0.07304   & 0.23075 \\ \hline 10^{14}& 8.22\times 10^{16} & 0.05988
& 0.05963        & 0.5947    & 0.11540 \\ \hline
\end{array}
\]
\caption{Two-loop results for the $SU(5)$ GUT scale $M_U$ and the `shifted' gauge coupling
values $a_i(M_U)$ in the case of two vector-like $Q+\bar Q$ quark pairs and
three $u^c+\bar u^c$ pairs. The corresponding decoupling scale $M_X$ is shown in the first column. }
\label{2LR}
\end{table}

We close our analysis with a remark concerning the gaugino masses.
The above GUT relations  and the threshold corrections induced by the fluxes discussed above
are expected to have also significant implications on the gaugino masses $M_i$ whose
magnitude at low energies is determined by the renormalization group equations
\be
\frac{d\,M_i}{d\,t} = \frac{b_i}{2\pi}\,a_i\,M_i
\ee
Combining with the RG equations for the gauge couplings $a_i$ we find that at
any scale $t=\ln \mu$ the following equation holds
\be
\frac{M_i(\mu)}{a_i(\mu)} = \frac{M_i(M_U)}{a_i(M_U)}
\ee
where $M_i(M_U), a_i(M_U)$ are the corresponding values at the unification scale. Since
in a unified gauge group the gaugino masses belong to the same multiplet, we expect
$M_i(M_U)=m_{1/2}$. This implies the following relation between the gaugino masses
irrespectively of the unification scale and the mass spectrum of the theory
\be
2\,\frac{M_3}{a_3}+3 \,\frac{M_2}{a_2}-5\,\frac{M_1}{a_1} = 0\label{GMR}
\ee
In Table(\ref{guagmass}) we give the predictions of the three masses for the cases discussed
previously. Since the gaugino masses play a decisive role in the calculation of
the various scalar masses, we note that these definite mass relations would have also a
 significant impact on the scalar spectrum of the model.
\begin{table}[!t]
\[
\begin{array}{|c|ccc|ccc|ccc|}
\hline
        &\multicolumn{3}{c}{m_{1/2}=300}&\multicolumn{3}{|c}{m_{1/2}=350}&\multicolumn{3}{|c|}{m_{1/2}=400}\\
\hline
M_X     &   M_1   &   M_2   &   M_3    &   M_1   &   M_2   &   M_3    &   M_1   &   M_2   &   M_3    \\
\hline
10^{11} &    48   &   100   &   300    &   56    &   117   &   347    &   64    &   134   &   394    \\
\hline
10^{12} &    62   &   128   &   392    &   72    &   149   &   453    &   83    &   171   &   515    \\
\hline
10^{13} &    76   &   156   &   482    &   89    &   182   &   559    &   102   &    208  &   634    \\
\hline
10^{14} &    90   &   183   &   572    &   106   &   214   &   663    &   121   &  245    &   752    \\
\hline
\end{array}
\]
\caption{The gaugino masses for three cases of $m_{1/2}$ and four possible $M_X$ scales. The extra matter
consists of two pairs of Q's and 3 pairs of u's. All values are in GeV units.}
\label{guagmass}
\end{table}

Up to now we have worked out in detail the effects of the fluxes
on the string scale  and the gaugino masses, however we have
ignored threshold corrections which arise when integrating out Kaluza-Klein(KK) modes.
Since the KK-scale may differ substantially from the decoupling scale in a
given model~\cite{Donagi:2008kj,Palti:2009}, it is important to give an estimate
of the effect these thresholds can have on the various RG depended quantities.
In the F-version of the $SU(5)$ model that we discuss here, we have assumed the existence
of chiral matter residing only on the curves defined by the 7-brane intersections. Therefore,
we may assume threshold corrections arising only from KK-modes of gauge fields propagating
in the bulk  and those of chiral matter of the $\Sigma_{10}$
and $\Sigma_{\bar 5}$ matter curves. Denoting with $\delta_i=\delta_i^g+\delta_i^c+\delta_i^h$
the gauge, chiral matter and Higgs KK-threshold contributions to the three gauge couplings ($i=1,2,3$)
respectively, we find that the modified unification scale $M_U'$ is given by
\begin{equation}
M_U'=e^{-\frac{2\pi}{\beta_x}\delta}M_U\label{newMU}
\end{equation}
where $M_U$ is given by (\ref{M_U}) and $\delta$ is the combination
\begin{equation}
\delta=\frac 53\delta_1-\delta_2-\frac 23\delta_3\label{dels}
\end{equation}
To estimate the effects of these thresholds on the unification
scale, we follow closely the analysis of~\cite{Donagi:2008kj}.
 The massive modes contributing to these thresholds constitute the spectrum of the
Dirac operator in the eight dimensional theory which in the compactified four-dimensional
theory decomposes to Dolbeault operators  of the corresponding holomorphic bundle $V$ with  representation
$R_V$, i.e., $\bar\partial :\Omega_S^{0,k}\otimes R_V\ra\Omega_S^{0,k+1}\otimes R_V,\;k=0,1$.
The quantities involved in the thresholds are related to
the eigenvalues of the corresponding Laplacian
$\Delta_k=\bar\partial\bar\partial^{\dagger}+\bar\partial^{\dagger}\bar\partial$ acting on
$\Omega_S^{0,k}\otimes R_V$ and can be
expressed in terms of the logarithm of the determinant
$\log \det'\Delta=-\zeta'_{\Delta}(0)$ ~\cite{Hortacsu:1979fg,Weisberger:1986qd}, where
the prime in $\det'$ means that the zero modes are excluded.

For the bulk gauge fields we may consider the decomposition (\ref{24bu}) and
denote  each representation $R_Y$ with respect to its hypercharge $Y=0,\pm 5/6$,
while for each group factor the corresponding contribution is
\begin{equation}
\delta_i^g=\frac 1{4\pi}\sum_Y2 Tr_{R_Y}(Q_i^2)K_Y\label{Thres}
\end{equation}
where
\begin{equation}
K_Y=2\log \det{'}\frac{\Delta_{0,R_V}}{M^2}-\log \det{'} \frac{\Delta_{1,R_V}}{M^2}
\end{equation}
It has been argued~\cite{Donagi:2008kj} that this type of thresholds can be expressed in terms of the Ray-Singer
torsion $T_{R_V}$~\cite{Ray:1973sb} modulo the $M^2$ dependence. The corrections may finally be written
\begin{equation}
\delta_i^g=\frac{b_i^g}{4\pi}\log\frac{M^2}{\mu^2}+\tilde\delta^g_i
\end{equation}
with $\tilde\delta^g_i=\frac{b^{(5/6)}_i}{2\pi}(T_{5/6}-T_0)$.
 To get an estimate of the order of corrections, following~\cite{Donagi:2008kj}
we choose a special case of line bundle ${\cal O}(n,-n)$ on $P^1\times P^1$
and apply for the case $n=1$, so that $L^5={\cal O}(1,-1)$. Using the
fact that $T_{{\cal O}(k)}=-\frac 12\zeta_k'(0)$ and the results of~\cite{Hortacsu:1979fg,Weisberger:1986qd}
we find
\begin{equation}
\tilde\delta^g=\frac{4}{\pi}(T_{{\cal O}(-1)}-T_{{\cal O}(0)})=-\frac 1{\pi}
\end{equation}
which implies a correction of the order $M_U'\sim 1.22\times M_U$ on the scale $M_U$.

To estimate the remaining threshold effects we first note that all  chiral and vector
like matter as well as the Higgs fields, are localized on the $\Sigma_{\bar 5}$ and $\Sigma_{10}$ curves.
Using the analogue of formula  (\ref{Thres}) we find that only the
$\Sigma_{10}$-thresholds contribute  to $M_U$ since for the $\Sigma_{\bar 5}$ curve
the threshold combination (\ref{dels}) is found to be zero.
Following the same reasoning with~\cite{Donagi:2008kj}, we consider the simplest
non-trivial case of $\Sigma_{10}$ being of genus one and  express the corrections
in terms of the torsion of flat line bundles $L_z$, as follows
\begin{equation}
\delta^{10}=\frac{1}{4\pi}(4T_{\cal O}-4T_{L_z})=\frac{1}{\pi}(T_{\cal O}-T_{L_z})
\end{equation}
Substituting in the exponent of (\ref{newMU})  and noting that the difference
$T_{\cal O}-T_{L_z}$ is positive for a large $z$-range~\cite{Donagi:2008kj},
we infer  that contributions from $\Sigma_{10}$  tend to abate the effect
of the thresholds on $M_U$ since as we can easily observe  this
correction works in the opposite direction of the gauge threshold discussed above.
Although there is no a priori reason that these two contributions are
equal in magnitude, we  may work out cases where  the total
effect is rather small compared to our previous calculations.

\section{Conclusions}

In the present work, we have discussed several phenomenological issues of the
 low-energy effective theory derived in the context of an F-theory $SU(5)$ GUT.
We have investigated the r\^ole of the Yukawa couplings generated from exotic
representations and other matter states --beyond those of the minimal
supersymmetric spectrum--   which are  usually present in F-theory constructions.
We have shown that the exotic color pairs $(\bar 3,2)_5+(3,\bar 2)_{-5}$
descending from the $SU(5)$ adjoint induce unacceptable mixing mass terms
while they lower dangerously the unification scale and they should
be diminished from the matter spectrum, in the way already described in \cite{Beasley:2008dc}.
Furthermore, we have suggested that a sensible way to evade the problem
of fast proton decay caused by the triplets found in the $5/\bar 5$ Higgs fields,
is to obtain a high enough  GUT breaking scale $M_U\gg 10^{16}$GeV.
Indeed,  taking advantage of the splitting of  gauge couplings~\cite{Blumenhagen:2008aw}
at $M_U$ --induced by the $SU(5)$-breaking $U(1)_Y$ flux-- and the presence of
appropriate types of exotic matter, we have shown that the GUT scale can take values
of the order $M_U\gtrsim 10^{17}$ GeV. This way, if the triplets  decouple at
the GUT scale and acquire a mass of the same order, they could suppress adequately the
relevant proton decay operators. We finally presented in brief the implications of
the gauge coupling splittings at $M_U$, on  the gaugino masses.

\vfill
\noindent
{\it Acknowledgements.} This work is partially
supported by the European Research and Training Network
MRTPN-CT-2006 035863-1 (UniverseNet) and the European Union ITN Programme
``UNILHC'' PITN-GA-2009-237920.
\\
\\
While this paper was reaching its final form, we noticed some recent
work\cite{Marsano:2009wr,Dudas:2009hu} where several similar issues
are also discussed.

\end{document}